\documentclass[12pt]{article}
\usepackage{geometry}
\usepackage{setspace}
\usepackage{caption}
\usepackage{times}
\usepackage{graphicx} 
\usepackage{wrapfig}
\usepackage{dcolumn}
\usepackage{bm}
\usepackage{subfigure}
\usepackage{amssymb,amsmath}
\usepackage{hyperref}
\usepackage[super,sort&compress]{natbib}

\title{A General Model for Static Contact Angles}
\author{Carlos E. Colosqui\\
Department of Mechanical Engineering, Stony Brook University\\
Department of Applied Mathematics \& Statistics, Stony Brook University\\
\texttt{carlos.colosqui@stonybrook.edu}}
\date{}

\begin{document}
\maketitle

\begin{abstract}
The problem of contact angle and hysteresis determination has direct implications for engineering applications of wetting, colloid and surface science. 
Significant technical challenges can arise under real-world operating conditions, because the static contact angle is strongly influenced by contamination at the liquid-solid and liquid-vapor interfaces, chemical aging over long times, and environmental variables such as relative humidity and temperature. Analytical models that account for these real-world effects are therefore highly desirable to guide the rational design of robust applications. This work proposes a unified and simple-to-use model that expands Young's local thermodynamic approach to consider surfaces with topographic features of general geometry and varying degrees of liquid infiltration. The unified model recovers classical wetting limits (Wenzel, Cassie-Baxter, and hemiwicking), accounts for observable intermediate states (e.g., impregnating Cassie), and identifies a new limiting state with potential realizability: a Cassie state accompanied by a precursor film, termed the Inverse Wenzel state.
\end{abstract}


\section{Introduction}
The determination of static contact angles on technologically relevant surfaces that exhibit micro- and nanoscale physical and chemical features remains a foundational problem in colloid and surface science.\cite{good1992contact,quere2002rough,meiron2004,huhtamaki2018surface,drelich2020contact} 
The problem of contact angle and hysteresis determination has direct implications for engineering applications ranging from surface wetting control,\cite{extrand2002model,xia2012anisotropic,Checco2014,aktar2020,al2022toward,shao2025adaptive} capillary infiltration in porous materials,\cite{gruener2009capillary,colosqui2011droplet,gruener2012anomalous,ceratti2015critical,colosqui2016,walls2016capillary,zhao2023anomalous} and particle adhesion at interfaces,\cite{kaz2011,colosqui2013,pack2015colloidal,yin2018dynamic,fu2024distinct,colosqui2024kinetic} to additive manufacturing.\cite{neukaufer2020investigation,mekhiel2021additive,ghosh2023diffusiophoresis}
The functionality and performance of devices across these applications are highly sensitive to how closely realized contact angles and hysteresis ranges agree with experimental and analytical predictions.
Significant technical challenges can arise under real-world operating conditions, because the static contact angle, is strongly influenced by surface cleanliness and contamination at the liquid-solid and liquid-vapor interfaces, by chemical aging processes over long times, and by environmental variables such as relative humidity and temperature.
Analytical models that account for these real-world effects are therefore highly desirable to guide the rational design of robust applications.

For an ideal surface that is perfectly flat and chemically homogeneous, and a volatile liquid in thermodynamic equilibrium with its vapor phase, the equilibrium contact angle is unique and given by Young's law,\cite{young1805}
\begin{equation}
\gamma_{\ell v}\cos\theta_Y = \gamma_{sv} - \gamma_{s \ell},
\label{eq:young}
\end{equation}
which relates the equilibrium (Young) contact angle $\theta_Y$ to the interfacial surface energies $\gamma_{s \ell}$, $\gamma_{\ell v}$, and $\gamma_{sv}$ of the solid-liquid, liquid-vapor, and solid-vapor interfaces, respectively. 
The interfacial surface energies in Eq.~\ref{eq:young} are conventionally treated as material properties uniquely determined by the particular surface and fluid chemistry.
Although simple tangential force-balance derivations of Eq.~\ref{eq:young} for a plane surface have known flaws, rigorous thermodynamic arguments have established the validity of Young's law for general geometric configurations.\cite{roura2004local,makkonen2016young} 
Moreover, recent studies indicate that Young's law remains applicable at unexpectedly small length scales, such as contact-line curvature and surface feature dimensions, down to roughly ten liquid molecule diameters.\cite{elliott2021surface,huang2022effects,teshima2022quantifying}

Engineered or natural surfaces with micro/nanoscale roughness or physical topography, exhibit a range of static contact angles that can significantly differ from the Young angle (Eq.~\ref{eq:young}) and are observed as the contact line quasi-statically recedes or advances. 
To account for the observed static contact angles and hysteresis phenomenon, two wetting models have been widely adopted: the Wenzel model\cite{wenzel1949surface}
\begin{equation}
\cos\theta_{W} = r\,\cos\theta_{Y},
\label{eq:wenzel}
\end{equation}
and the Cassie-Baxter model\cite{cassie1944} 
\begin{equation}
\cos\theta_{CB} = \varphi_{S}\,\cos\theta_{Y} - \big(1-\varphi_{S}\big),
\label{eq:cassie-baxter}
\end{equation}
for the apparent equilibrium contact angles $\theta_{W}$ and $\theta_{CB}$ corresponding to each state, $r\ge 1$ is ratio of the actual wetted solid area to its projected area, and $\varphi_S$ is the planar solid area fraction in contact with the wetting liquid phase.
It is important to note that the Wenzel and Cassie-Baxter models in Eqs.~\ref{eq:wenzel}--\ref{eq:cassie-baxter} represent two limiting wetting states: full liquid impregnation of the surface topography (Wenzel state) and strictly zero liquid infiltration beneath the liquid phase (Cassie-Baxter state).
Notably, intermediate wetting states with partial and localized infiltration of the surface topography are realized under typical ambient conditions, and these commonly correspond to metastable configurations separated by large but finite free-energy barriers.\cite{whyman2011,mihalis2012,giacomello2012,mihalis2013,moradi2014contact,bormashenko2015progress}
In particular, certain combinations of micro/nanoscale topography and low intrinsic Young angle, promote stable wetting states with the infiltration or liquid ahead of the contact line, through the phenomenon of hemiwicking.\cite{bico2002wetting,chen2019toward,patankar2021,panter2023rough}

Prior work has focused on developing \textit{unified models} that bridge the pure Wenzel and Cassie-Baxter extremes and extend these classical models.\cite{whyman2008rigorous,raj2012unified,rohrs2019wetting}
These efforts employ free energy minimization to account for multiple wetting states and assess their thermodynamic stability, for specific wetting configurations and surface geometries such as  a macroscale sessile droplet on much smaller micropillar arrays.
Despite the significant progress on the fundamental and applied aspects, open questions remain, namely how to predict static contact angles and the hysteresis range from a knowledge of topographic parameters, surface chemistry aging, and ambient conditions affecting the degree of liquid infiltration on the surface below and ahead of the wetting liquid phase.
In an effort to address this, this work proposes a general and simple-to-use model, expanding Young's local thermodynamic approach to consider surfaces with topographic features of general geometry and varying degrees of liquid infiltration.
The introduced unified model recovers classical wetting limits (Wenzel, Cassie-Baxter, and hemiwicking states), accounts for observable intermediate states (e.g., impregnating Cassie), and further identifies a new limiting state with potential realizability: a Cassie state accompanied by a precursor film, here termed the Inverse Wenzel state.

The introduced model aims to predict the static contact angle $\theta$ from the substrate and interfaces chemistry, characterized by the Young contact angle $\theta_Y$, and geometric parameters of the surface topography, such as the top, lateral, and basal surface areas, for generalized wetting conditions prescribed by arbitrary degrees of liquid infiltration beneath the bulk liquid and ambient phases.
It is worth noting that the liquid infiltration fractions on the surface topography adjacent to the contact line, treated as known inputs in this derivation, result from kinetically trapped (metastable) or thermodynamically stable states that can coexist. 
These infiltration fractions may or may not be realized under the specific physicochemical conditions and/or dynamic constraints in the actual system studied, as in the case of the classical Wenzel and Cassie-Baxter models (Eqs.~\ref{eq:wenzel}--\ref{eq:cassie-baxter}). 

\begin{figure*}[h!]
\begin{center}
\includegraphics[width=1\textwidth]{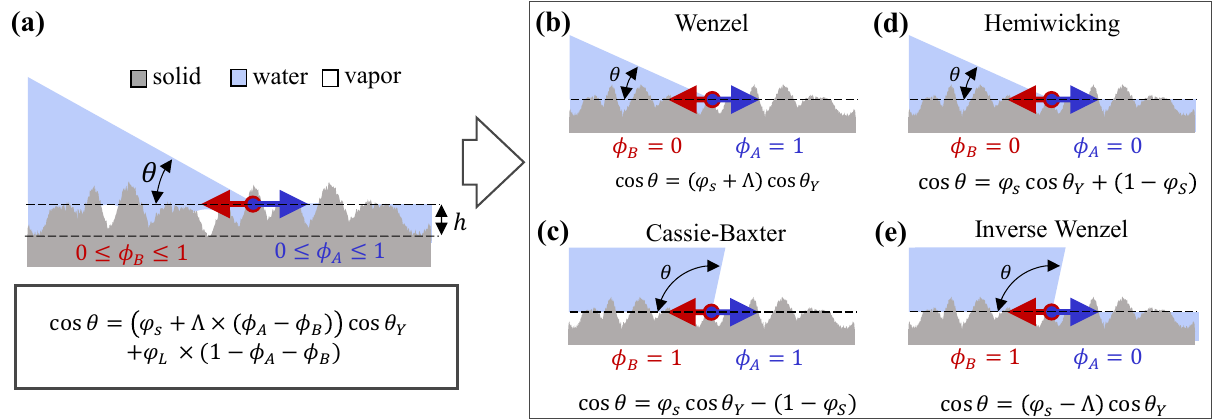}
\end{center}
\caption{Unified model for static contact angles on a surface with micro/nanoscale surface topography. 
(a) Generalized wetting state and unified model for arbitrary degrees of liquid infiltration in the bulk side ($0\le\phi_B\le 1$) and ambient side ($0\le\phi_A\le 1$) of the contact line. 
(b) Wenzel state corresponding to full liquid infiltration under the liquid phase ($\phi_B=0$) and a perfectly dry surface under the ambient phase ($\phi_A=1$).   
(c) Cassie-Baxter state corresponding to strictly no liquid infiltration under the liquid and ambient phases ($\phi_B=\phi_A=1$).
(d) Hemiwicking state corresponding to full liquid infiltration under the liquid and ambient phases ($\phi_B=\phi_A=0$).
(e) Inverse Wenzel state corresponding to strictly no liquid infiltration under the liquid phase ($\phi_B=1$) and full liquid infiltration under the ambient phase ($\phi_A=0$).   
\label{fig:1}}
\end{figure*}

As illustrated in Fig.~\ref{fig:1}a, the liquid phase lies on the Bulk (B) side of the contact line and the vapor phase on the Ambient (A) side.   
We consider a chemically homogeneous substrate with (random or regular) three-dimensional micro- and nanoscale topographic features (Fig.~\ref{fig:1}a) producing a full surface area $S\ge A$, larger than the projected surface area $A$. 
\paragraph{Geometric parameters.} The full surface area is generally decomposed as $S=S_{\text{top}}+S_{\text{lat+bot}}$, where $S_{\text{top}}\!\ge 0$ is the planar area of feature tops, if present at all, and $S_{\text{lat+bot}}\!>0$ denotes the lateral plus basal area generated by the topography. 
We thus define the solid-top flat area fraction $\varphi_S=S_{\text{top}}/A$ and the lateral-basal area fraction $\Lambda = S_{\text{lat+bot}}/A$.
With this notation, the Wenzel roughness ratio in Eq.~\ref{eq:wenzel} is $r = S/A = \varphi_S + \Lambda$.
We will define $\varphi_L(h)=A_{\ell v}/A$ as the area fraction that would correspond to liquid-vapor interface if the surface topography were fully infiltrated by liquid up to the reference height $h$ of micro- or nanoscale dimensions. 
We will additionally assume that the reference liquid height $h$ within the infiltrated topography sets the local contact line position and is nearly constant within small distances $\lambda\sim h$ from the contact line.
Note that, under assumptions leading to the Cassie-Baxter model (Eq.~\ref{eq:cassie-baxter}), we have $\varphi_L = 1-\varphi_S$ but this equality is strictly valid when all topographic features have approximately the same height $h_r$ and the liquid-vapor interface is perfectly flat and coincides with the reference level $h=h_r$; see Ref.~\citenum{milne2012cassie} for a detailed discussion.
For random self-affine micro/nanoscale topography with tall peaks of small lateral extent protruding above the r.m.s. height, one may approximate $\varphi_L \simeq 1-\varphi_S$; an error on the order of roughly $15\%$ can be estimated for Gaussian height statistics.

\paragraph{Generalized wetting conditions.} Deviating from the conventional treatment that leads to the classical Wenzel and Cassie-Baxter states, we consider that in general wetting and infiltration of the surface topography may be partial and nonuniform; near the contact line, localized air/vapor voids may exist beneath the liquid phase, while localized liquid infiltration by wicking and condensation may occur through the texture on the ambient side exposed to the ambient vapor phase (see Fig.~\ref{fig:1}a).
Hence, let $\phi_B \in [0,1]$ denote the averaged fraction of surface features occupied by air/vapor in the bulk side of the contact line and $\phi_A\in [0,1]$ denote the averaged fraction of surface features filled with air/vapor in the ambient side of the contact line.
The limit condition with zero liquid-vapor interfacial area $S_{\ell v}=0$ on the bulk and ambient side thus correspond to $\phi_{B}=0$ and $\phi_{A}=1$; i.e., the Wenzel state (Fig.~\ref{fig:1}b).
Similarly, the limit with $\phi_{B}=1$ and $\phi_{A}=1$ corresponds to the case that the liquid-vapor interface is $S_{\ell v}=\varphi_L A$; i.e., the Cassie-Baxter state (Fig.~\ref{fig:1}c).

\paragraph{Free-energy stationarity.}
Following a conventional thermodynamic argument, we will consider that receding or advancing displacements of the contact line, under constant temperature, volume, and chemical equilibrium, produce a change in the free energy (or grand potential)\cite{bico2002wetting,colosqui2017} 
$dF=\gamma_{\ell v} \cos\theta dA + \gamma_{s\ell} dS_{s\ell} +\gamma_{sv} dS_{sv}
+\gamma_{\ell v} dS_{\ell v}$,
where $\gamma_{s\ell}$, $\gamma_{sv}$, and $\gamma_{\ell v}$ are the solid-liquid, solid-vapor, and liquid-vapor interfacial surface energies, and $S_{s\ell}$, $S_{sv}$, and $S_{\ell v}$ are the corresponding interfacial surface areas.
Based on the geometric considerations above, a local displacement of the contact line over a projected area differential $\Delta A$, produces the following interfacial areas change:
\begin{align}
\Delta S_{s\ell} &= \Big(\,\varphi_S \;+\; \Lambda(\phi_A-\phi_B)\,\Big) \Delta A, \label{eq:dSls}\\[2pt]
\Delta S_{sv} &= \Big(\,-\varphi_S \;+\; \Lambda(\phi_A-\phi_B)\,\Big) \Delta A, \label{eq:dSsv}\\[2pt]
\Delta S_{\ell v} &= -\,\Delta A\,\varphi_L\,(1-\phi_A-\phi_B).\label{eq:dSlv}
\end{align}
By imposing free-energy stationarity $dF=0$, invoking the Young contact angle definition in Eq.~\ref{eq:young}, and introducing the surface area changes in Eqs.~\ref{eq:dSls}--\ref{eq:dSsv}, one obtains that the contact angle for static equilibrium must satisfy 
\begin{equation}
\cos\theta \;=\; \cos\theta_Y \,\Big(\varphi_S + \Lambda(\phi_A-\phi_B)\Big) \;+\; \varphi_L\,\big(1-\phi_A-\phi_B\big). \;
\label{eq:master}
\end{equation}

The general model for the local contact angle derived in Eq.~\ref{eq:master} recovers the limiting behaviors predicted by previous well-established models, by making the conventional assumption that $\varphi_L+\varphi_S=1$: 
in the Wenzel state with $\phi_B=0$ and $\phi_A=1$, Eq.~\ref{eq:master} reduces to Eq.~\ref{eq:wenzel} (Fig.~\ref{fig:1}b);
in the Cassie-Baxter state for which $\phi_B=1$ and $\phi_A=1$ and the surface topography is fully filled with air/vapor on both the bulk and ambient sides, Eq.~\ref{eq:master} yields Eq.~\ref{eq:cassie-baxter} (Fig.~\ref{fig:1}c); 
and in the hemiwicking configuration (Fig.~\ref{fig:1}d), for which the surface topography is fully infiltrated by liquid on both sides adjacent to the contact line ($\phi_B=0$ and $\phi_A=0$), Eq.~\eqref{eq:master} simplifies to
$\cos\theta=\varphi_S\cos\theta_Y+(1-\varphi_S)$.\cite{bico2002wetting}
In addition, Eq.~\ref{eq:master} predicts a fourth limiting state, here termed the Inverse Wenzel state (Fig.~\ref{fig:1}e), for which $\phi_B=1$ and $\phi_A=0$ and Eq.~\ref{eq:master} gives
\begin{equation}
\cos\theta \;=\; \cos\theta_Y (\varphi_S - \Lambda).
\label{eq:iw}
\end{equation}
This inverted state, which can, for example, yield large local contact angles $\theta\gtrsim 90^\circ$ on a hydrophilic substrate with $\theta_Y<90^\circ$, can be realized when the surface topography traps a localized metastable vapor void next to a region highly infiltrated by liquid. 

\section{Results and Discussion}
This section presents analytical results from Eq.~\ref{eq:master} for a wide range of physically meaningful Young contact angles~$\theta_Y$ and different surface topographies characterized by their lateral-basal area fraction $\Lambda$ and top-solid flat area fraction $\varphi_S$. The analysis covers the full range of topography infiltration conditions, characterized by the liquid-vapor area fractions on the bulk and ambient sides adjacent to the contact line.
For simplicity, the results reported here assume~$\varphi_L \simeq 1 - \varphi_S$.

\subsection{Sharp random topography}
%
\begin{figure*}[h!]
\begin{center}
\includegraphics[width=1\textwidth]{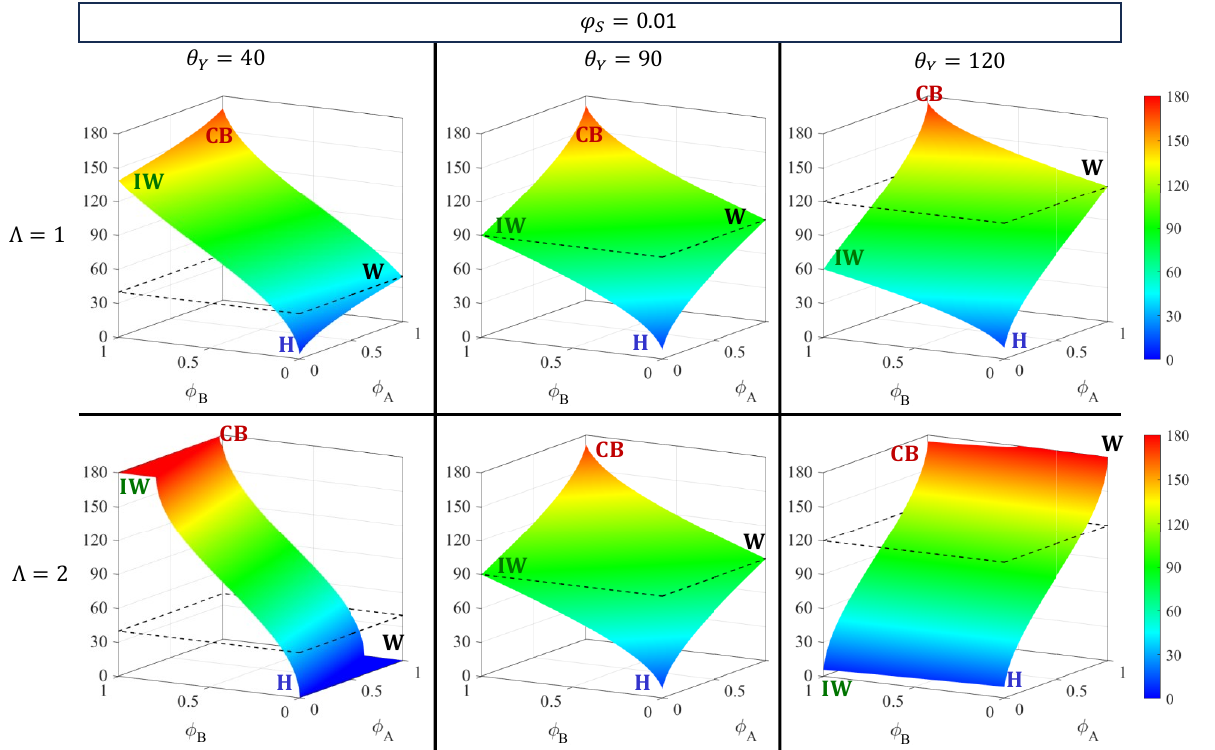}
\end{center}
\vskip -15 pt
\caption{Sharp random topography. Local static contact angle~$\theta$~ predicted by Eq.~\ref{eq:master} as a function of vapor fractions~$\phi_B$~(bulk) and~$\phi_A$~(ambient) adjacent to the contact line, for three Young angles: $\theta_Y = 40^\circ,\,90^\circ,\,\&\,120^\circ$. 
The results correspond to surface topography with a small solid-top flat area fraction $\varphi_S = 0.01$ and for two different lateral-basal area fractions $\Lambda = 1$ (top row) and $\Lambda = 2$ (bottom row). Wetting states: (W) Wenzel, (CB) Cassie-Baxter, (IW) Inverse Wenzel, and (H) Hemiwicked.
\label{fig:2}}
\end{figure*}
%
The results in Fig.~\ref{fig:2} correspond to surfaces with sharp topographic features, with a small solid-top flat area fraction $\varphi_S = 0.01$ and moderate-to-large lateral-basal area fractions $\Lambda = 1$ and $\Lambda = 2$.
The combination of topographic parameters in Fig.~\ref{fig:2} represents, for example, common glass surfaces with natural micro/nanoscale roughness\cite{colosqui2016,jose2018physical,zhao2023anomalous} or engineered nanostructures with random nanomaterial deposition.\cite{nandyala2020design,wang2021glass}
The analytical results in Fig.~\ref{fig:2} show that a wide range of static contact angles, which deviate largely from the intrinsic Young contact angle, is feasible for this type of surface topography.
For sufficiently ``tall'' topographic features giving large lateral-basal area fractions $\Lambda \ge 2$, robust superhydrophilicity with $\theta \to 0$ can be realized for $\phi_B = 0$ and $0 \le \phi_A \le 1$ on wettable substrates with $\theta \lesssim 40^\circ$.
This indicates that superwetting with vanishing hysteresis can be attained regardless of the conditions on the surface exposed to the ambient phase.
Notably, large lateral-basal area fractions $\Lambda \ge 2$ in a hydrophilic substrate can promote superhydrophobic states with $\theta \ge 150^\circ$ if the liquid phase is nearly fully suspended under topographically trapped air ($\phi_B \ge 0.9$), for arbitrary wetting conditions on the ambient side ($0 \le \phi_A \le 1$).
This surprising result corresponds to the so-called Inverse Wenzel state identified in Fig.~\ref{fig:1}e.

\subsection{Engineered pillared structures}
\begin{figure*}[h!]
\begin{center}
\includegraphics[width=1\textwidth]{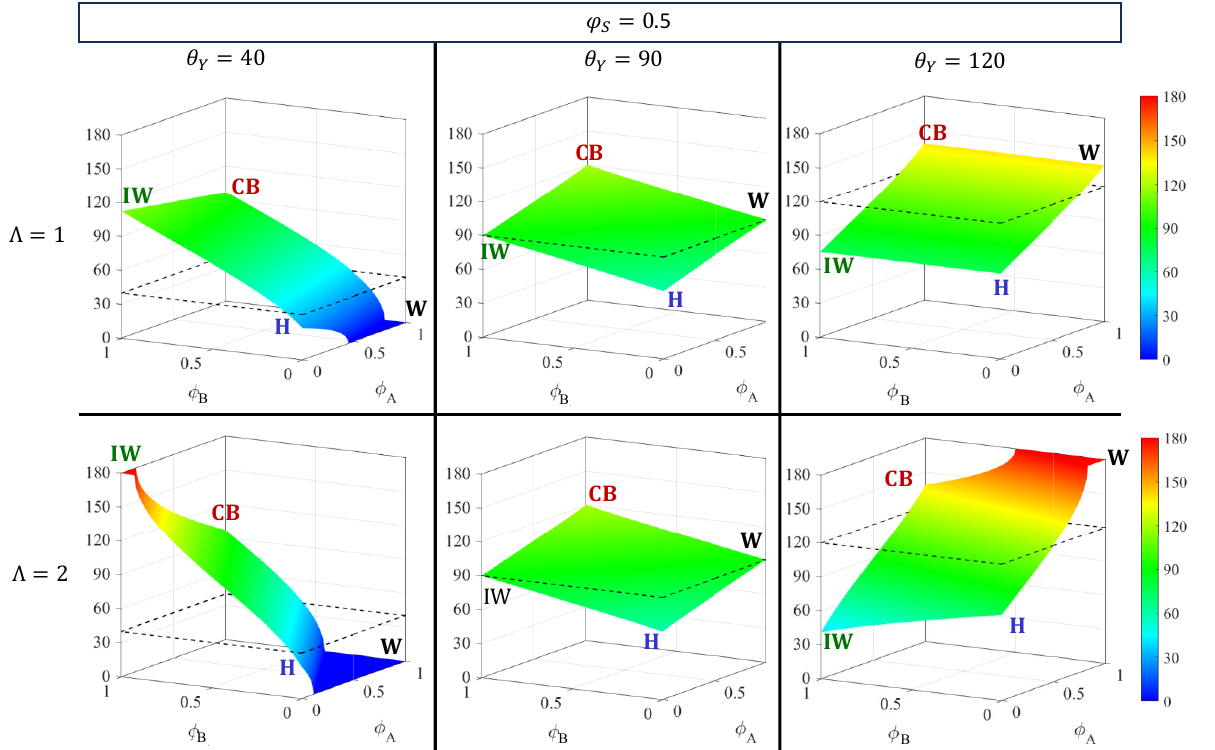}
\end{center}
\vskip -15 pt
\caption{Engineered pillared structures. Local static contact angle~$\theta$~ predicted by Eq.~\ref{eq:master} as a function of vapor fractions~$\phi_B$~(bulk) and~$\phi_A$~(ambient) adjacent to the contact line, for three Young angles: $\theta_Y = 40^\circ,\,90^\circ,\,\&\,120^\circ$. 
The results correspond to surfaces with sharp topographic features, with a small solid-top flat area fraction $\varphi_S = 0.5$ and for two different lateral-basal area fractions $\Lambda = 1$ (top row) and $\Lambda = 2$ (bottom row). Wetting states: (W) Wenzel, (CB) Cassie-Baxter, (IW) Inverse Wenzel, and (H) Hemiwicked.
\label{fig:3}}
\end{figure*}
%
The static contact angle predictions in Fig.~\ref{fig:3} correspond to surface topographies with a moderately large top-solid area fraction, $\varphi_S = 0.5$, and lateral-basal area fractions, $\Lambda = 1$ and $\Lambda = 2$.
This type of surface topography is typically attained through engineered surface structures of micro- or nanoscale dimensions, featuring flat top surface areas with controlled dimensions and lateral-basal surface areas determined by the feature height and period (see, for example, Refs.~\citenum{checco2014collapse,aktar2020,al2022toward}).
In this case, analytical predictions show less sensitivity to variations in the wetting conditions on the surface exposed to the liquid bulk and ambient vapor phase, resulting in relatively moderate hysteresis ranges for Young angles, $\theta_Y$, between 60$^\circ$ and 120$^\circ$, and shallow surface features for which $\Lambda \simeq 1$ (cf. Fig.~\ref{fig:3}).
As shown in Fig.~\ref{fig:3}, surfaces with a moderately large top-solid area fraction, $\varphi_S = 0.5$, require low Young angles, $\theta_Y \lesssim 40^\circ$, and tall structures with short periods that yield large lateral-basal area fractions,
$\Lambda \gtrsim 2$, to produce robust superhydrophilicity for varying conditions on the surface side exposed to the ambient phase.
The realizability of large contact angles, $\theta \geq 150^\circ$, for hydrophilic substrates with $\theta_Y \simeq 40^\circ$ is confined to a very narrow range of conditions ($\phi_B > 0.9$ and $\phi_A < 0.1$), for which the so-called Inverse Wenzel state can exist.
In addition, sticky superhydrophobicity with $\theta \ge 150^\circ$ can be attained on this type of surface topography (cf. Fig.~\ref{fig:3}) when the substrate is highly hydrophobic ($\theta_Y \geq 120^\circ$) and the liquid bulk infiltrates the topography underneath while the ambient side is partially wetted ($\phi_B \le 0.5$ and $\phi_A \ge 0.9$).    

\subsection{Micro/nanocavity arrays}
%
\begin{figure*}[h!]
\begin{center}
\includegraphics[width=1\textwidth]{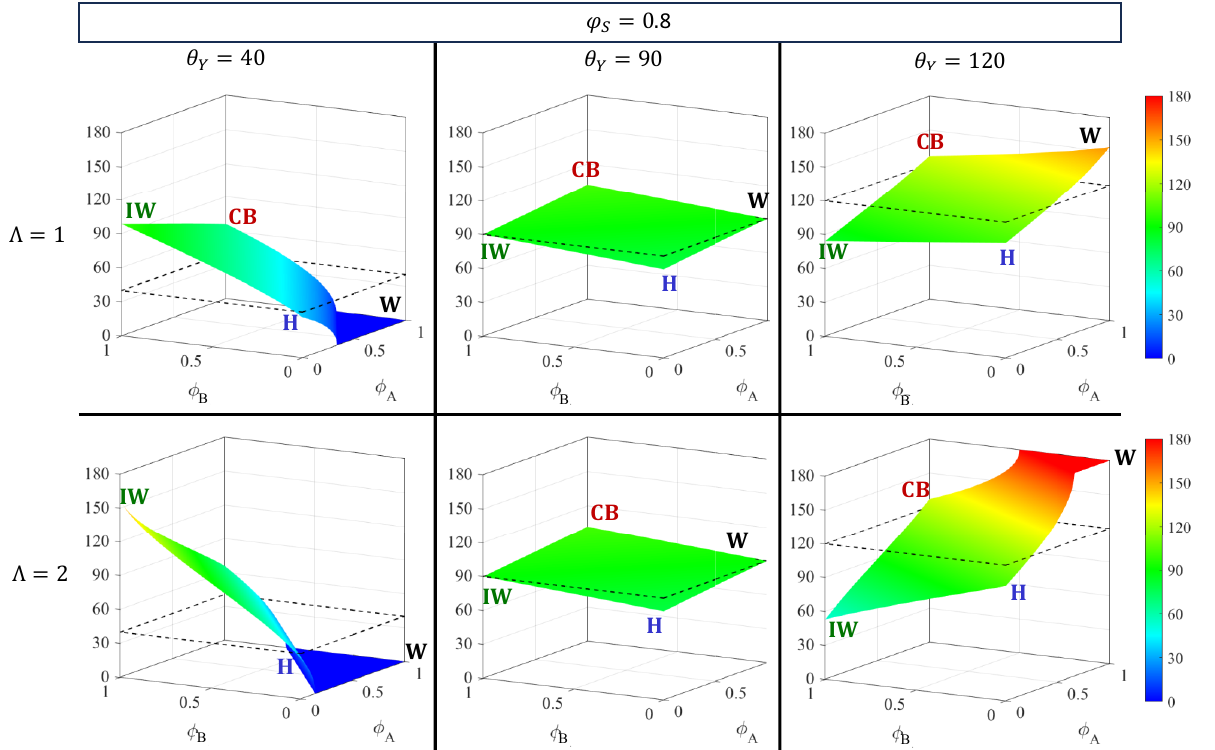}
\end{center}
\vskip -15 pt
\caption{Micro/nanocavity arrays. Local static contact angle~$\theta$~ predicted by Eq.~\ref{eq:master} as a function of vapor fractions~$\phi_B$~(bulk) and~$\phi_A$~(ambient) adjacent to the contact line, for three Young angles: $\theta_Y = 40^\circ,\,90^\circ,\,\&\,120^\circ$. 
The results correspond to surfaces with sharp topographic features, with a small solid-top flat area fraction $\varphi_S = 0.8$ and for two different lateral-basal area fractions $\Lambda = 1$ (top row) and $\Lambda = 2$ (bottom row). Wetting states: (W) Wenzel, (CB) Cassie-Baxter, (IW) Inverse Wenzel, and (H) Hemiwicked.
\label{fig:4}}
\end{figure*}
%
The final cases analyzed (cf. Fig.~\ref{fig:4}) correspond to a large top-solid flat area fraction, $\varphi_S = 0.8$, and lateral-basal area fractions $\Lambda = 1$ and $\Lambda = 2$.
This type of surface topography can be attained, for example, by patterning a smooth and flat surface with arrays of cavities with controllable micro- or nanoscale dimensions (e.g., Refs.~\citenum{checco2010,gang2013,sogaard2014study}).
Note that if the cavities are not interconnected, capillary condensation and/or direct prewetting of the surface, rather than hemiwicking, are feasible mechanisms for infiltrating the surface topography.
For this type of topography, robust superhydrophilicity with $\theta \to 0$ can be realized for $\phi_A \gtrsim 0.5$ and $\phi_B \lesssim 0.5$ in the case of large lateral-basal areas with $\Lambda > 1$ and hydrophilic substrates with $\theta_Y \lesssim 40^\circ$.
Superhydrophobicity is more difficult to achieve with this type of surface topography, and $\theta \ge 150^\circ$ is predicted only for highly hydrophobic substrates when $\phi_A \gtrsim 0.8$ and $\phi_B \lesssim 0.5$, which represent a rather narrow range of conditions.

\section{Conclusions}
This work introduces a unified model for static contact angles that extends Young's local thermodynamic framework to surfaces with micro- or nanoscale topographic features and nonuniform liquid infiltration near the contact line. 
The model employs as input parameters the solid-top flat area fraction $\varphi_S$ and the lateral plus basal area $\Lambda$, and considers side-specific liquid infiltration under the liquid bulk and ambient phase through the vapor fractions $\phi_A\in[0,1]$ and $\phi_B\in[0,1]$. 
The unified model recovers the Wenzel, Cassie-Baxter, and hemiwicking limits, and it predicts a fourth limiting configuration of practical interest, an Inverse Wenzel state with a Cassie-like core and a uniform liquid film under the ambient phase. 
Because all inputs can be obtained from surface measurements and standard force tensiometer data, the model can link geometric parameters, liquid infiltration degree, and static contact angles, and it yields clear bounds for advancing and receding states and thus for static hysteresis.

Analytical predictions from the unified model provide direct design guidance for contact angle control under varying interfacial and environmental conditions.
For sharp random surface features with very small top-solid flat area fractions $\varphi_S<0.1$ and large lateral-basal areas $\Lambda>1$, the model predicts robust superhydrophilicity with $\theta \to 0$ on intrinsically wettable substrates ($\theta_Y<90^\circ$) when the bulk side is liquid filled ($\phi_B = 0$), for arbitrary ambient-side conditions ($0\le\phi_A\le 1$). 
Notably, the same class of sharp topographic features can also support large contact angles $\theta \gtrsim 150^\circ$ on hydrophilic substrates through the Inverse Wenzel state if the liquid is nearly fully suspended over kinetically or physically trapped vapor on the bulk side ($\phi_B \gtrsim 0.9$), which highlights how large lateral areas $\Lambda>1$ can amplify variations of the attainable static contact angle $\theta$ and produce extremely large hysteresis ranges. 
For engineered pillar arrays with moderate $\varphi_S \sim 0.5$, sensitivity to local liquid/vapor infiltration $(\phi_A,\phi_B)$ and thus ambient conditions is reduced, and achieving superhydrophilicity requires tall features ($\Lambda >1$) and low Young contact angles, while superhydrophobicity appears only for very large Young contact angles $\theta_Y\ge 120^\circ$ (e.g., for silanized substrates) within a narrow window of side-specific vapor fractions $(\phi_A\gtrsim 0.9,\phi_B<0.5)$. 
For surfaces with large $\varphi_S \sim 0.8$, such as those produced by micro/nanoscale cavity arrays, superhydrophilicity is robustly attainable for hydrophilic substrates when $\Lambda \ge 1$ and the ambient side is partially infiltrated by liquid, whereas attaining high contact angles is limited to intrinsically hydrophobic surfaces and moderately narrow range of vapor/liquid infiltration fractions $(\phi_A\gtrsim 0.8,\phi_B\lesssim 0.5)$.

The model predictions also help account for the slow near-equilibrium relaxation that arises from transitions between numerous metastable states when a contact line moves over surfaces with random nanoscale topographies. 
In droplet spreading, capillary imbibition, and nanoparticle adsorption, relaxation proceeds through a sequence of quasi-static configurations where the contact line remains locally pinned while the local static contact angle can vary over a wide range of values. It is therefore possible that both the geometry of topographic features and environmental conditions controlling local variations in $\phi_A$ and $\phi_B$ prescribe the relaxation rates documented in prior works.~\cite{kaz2011,colosqui2013,colosqui2015,colosqui2016,colosqui2019diffusion,jose2018physical,nandyala2020design,wang2021glass,zhao2023anomalous}

The theoretical framework presented in this work suggests concrete directions for further theoretical developments and experimental verification. 
Future developments of the presented unified model may include a thermodynamic closure for $\phi_A$ and $\phi_B$, by considering the Laplace, Kelvin, and disjoining pressure, along with capillary entry thresholds determined from topographic analysis (e.g., AFM, SEM, or profilometry); thereby making the vapor fractions $(\phi_A, \phi_B)$ predictable state variables rather than free model parameters.
Additionally, the unified model can be readily applied for complex surface topographies, with overhangs or hierarchical features, for which $\varphi_L+\varphi_S>1$. In such cases, $\varphi_L$ should be obtained from the level set area at an appropriate reference height or from three-dimensional surface reconstructions. The unified model in Eq.~\ref{eq:master} is generally  valid when the properly determined values of $\varphi_L$ and $\Lambda$ are used.
Force-displacement measurements with the Wilhelmy plate method at low displacement rates can be used to determine advancing and receding contact angles and infer $(\phi_A,\phi_B)$ along the contact line path. This type of experimental analysis could validate and refine analytical predictions for the static contact angle and hysteresis range under different environmental conditions, identify transitions between limiting and intermediate states, and guide the potential realization of the Inverse Wenzel configuration.

\section*{Acknowledgments}
The author thanks Prof. H.A. Stone for valuable conversations.
This work was supported by the National Science Foundation under award CBET-2417797.
The theoretical model development for this work was supported by the Center for Mesoscale Transport Properties, an Energy Frontier Research Center funded by the U.S. Department of Energy, Office of Science, Basic Energy Sciences, under award DE-SC0012673.
%


\end{document}